\documentclass{article}

\usepackage{spconf}

\usepackage{amsmath,graphicx}
\usepackage{amssymb}
\usepackage{latexsym}

\usepackage{algorithm}
\usepackage{algpseudocode}
\algblockdefx{ForAllP}{EndForAllP}[1]%
  {\textbf{for all }#1 \textbf{do in parallel}}%
  {\textbf{end for}} 
\newcommand\blfootnote[1]{%
  \begingroup
  \renewcommand\thefootnote{}\footnote{#1}%
  \addtocounter{footnote}{-1}%
  \endgroup
}
\usepackage[hyphens]{url}
\usepackage[colorlinks,urlcolor=blue]{hyperref}

\usepackage{xcolor}

\begin{document}







\title{Dealing with heterogeneous 3D MR Knee images: A Federated Few-Shot Learning method with dual knowledge distillation} 

%

\name{\parbox{\linewidth}{\centering{Xiaoxiao He$^{1}$, Chaowei Tan$^{1}$, Bo Liu$^{3}$, Liping Si$^{4,5}$, Weiwu Yao$^{4}$, Liang Zhao$^{6}$, Di Liu$^{1}$, Qilong Zhangli$^{1}$, Qi Chang$^{1}$, Kang Li$^{2}$ and Dimitris N. Metaxas$^{1}$}}}
\address{$^{1}$Department of Computer Science, Rutgers University, USA\\
$^{2}$West China Biomedical Big Data Center, Sichuan University West China Hospital, China\\
$^{3}$Walmart Global Tech, USA\\
$^{4}$Department of Imaging, Tongren Hospital, Shanghai Jiao Tong University School of Medicine, China\\
$^{5}$Department of Radiology, Zhongshan Hospital, Fudan University, Shanghai, China.\\
$^{6}$SenseTime Research, China
}
\maketitle
\begin{abstract}
Federated Learning has gained popularity among medical institutions since it enables collaborative training between clients (e.g., hospitals) without aggregating data. However, due to the high cost associated with creating annotations, especially for large 3D image datasets, clinical institutions do not have enough supervised data for training locally. Thus, the performance of the collaborative model is subpar under limited supervision. 
On the other hand, large institutions have the resources to compile data repositories with high-resolution images and labels. Therefore, individual clients can utilize the knowledge acquired in the public data repositories to mitigate the shortage of private annotated images. In this paper, we propose a federated few-shot learning method with dual knowledge distillation. This method allows joint training with limited annotations across clients without jeopardizing privacy. The supervised learning of the proposed method extracts features from limited labeled data in each client, while the unsupervised data is used to distill both feature and response-based knowledge from a national data repository to further improve the accuracy of the collaborative model and reduce the communication cost. Extensive evaluations are conducted on 3D magnetic resonance knee images from a private clinical dataset. Our proposed method shows superior performance and less training time than other semi-supervised federated learning methods.
\blfootnote{Codes and additional visualization results are available at \url{https://github.com/hexiaoxiao-cs/fedml-knee}.}
\end{abstract}


\section{Introduction}
\label{sec:intro}
\begin{figure}[t]
    \centering    \includegraphics[width=0.90\linewidth]{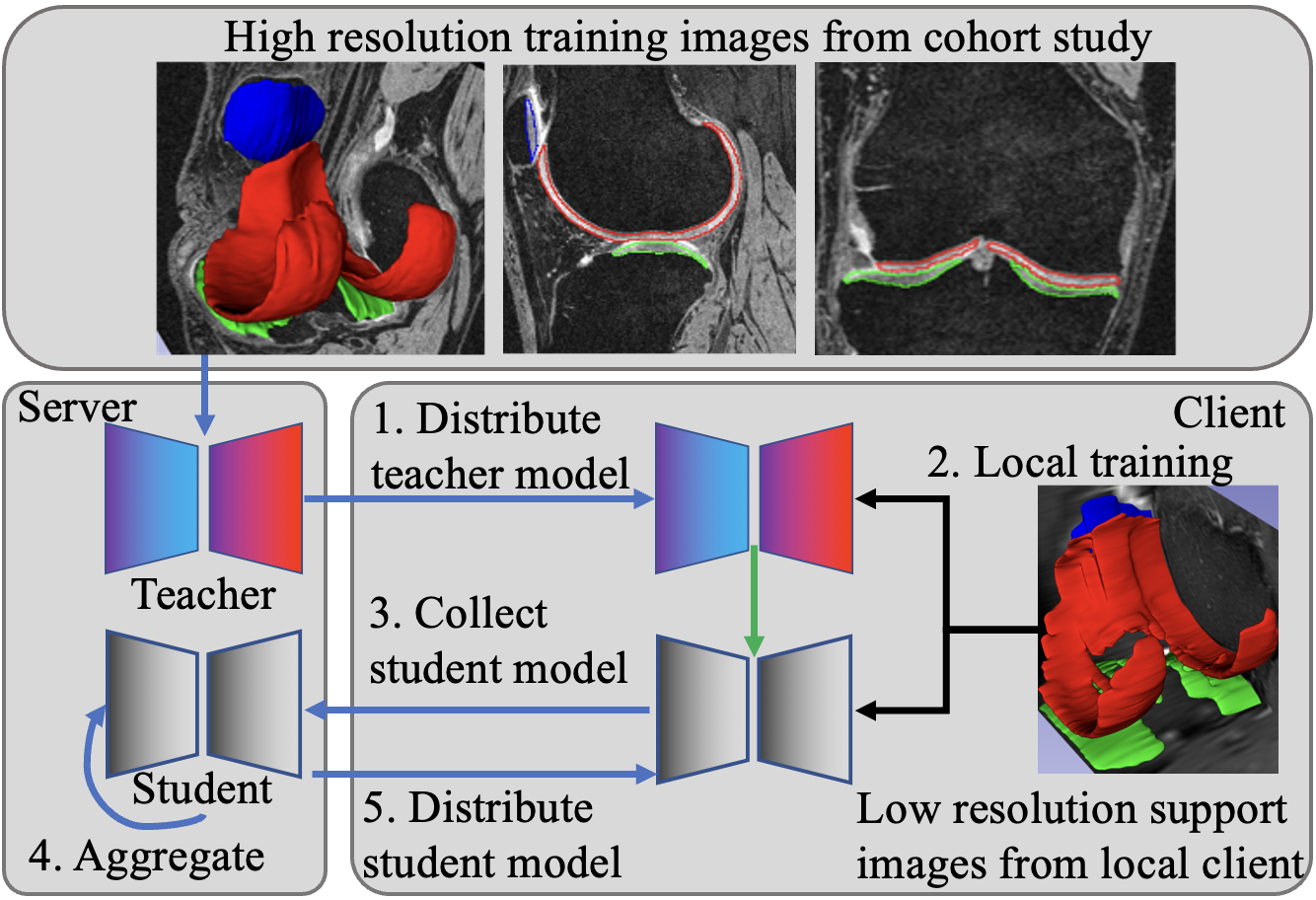}
    \caption{Overview of the proposed federated few-shot learning framework. We adopted the FL workflow from \cite{FedAVG}. The dual knowledge acquired from Osteoarthritis Initiative (OAI) training repository (top) is distributed to each client during initialization. Each client trains with local low-resolution support images (bottom) and returns the trained student model to the server. Both OAI and local knee MR data contain red, green, and blue annotations for femoral cartilage (FC), tibial cartilage (TC), and patellar cartilage (PC).}
    \label{fig:overview}
    \vspace{-0.5cm}
\end{figure}
Federated Learning (FL) \cite{FedAVG} was introduced as a privacy-aware framework that utilizes isolated datasets without aggregating data into one location. It uses a client-server architecture where each client performs training on the local dataset, and the server aggregates models submitted by clients as indicated in Fig.~\ref{fig:overview}. This framework is suitable for medical analysis applications \cite{Cov19Dayan,Dou:2021wd}, since it enables the cooperation between medical institutions to train the same model while preserving privacy. Although FL is beneficial for ensuring privacy during collaboration, each client model still needs to be accurate for the joint model to become useful in clinical applications \cite{chang2022deeprecon, liu2022transfusion, zhangli2022region}. However, medical institutions usually lack labeled data due to insufficient annotation resources, especially for 3D images. For example, our private dataset only contains 20 labeled images per client. 
Previously, Zhang et al. \cite{SSFL} tackled the problem of lacking annotated data by performing semi-supervised training on all clients by maximizing the expectation of pseudo labels among unlabeled data. Yang et al. \cite{FedSemiSupervised} employed the FixMatch consistency regularization among unlabeled data for better training. However, these semi-supervised methods struggle to produce an accurate segmentation network due to limited annotations. They also require more communication between clients and the server, which is challenging for institutions in remote areas. Therefore, more supervision during training is needed to reduce data traffic and improve accuracy.

Few-shot learning (FSL) utilizes prior knowledge acquired from a training set with sufficient annotations to guide training on the support set with limited labels for improving the accuracy of the client model. 
In our federated FSL approach, the support set located on each client utilizes our private dataset with limited supervision. The training set on the server uses the OAI repository that is heterogeneous in resolution and imaging parameters, while containing significantly more images with annotations compared to the support set. 
Therefore, high-quality OAI data can be used to improve the accuracy of clients on their local datasets. Instead of distributing the data repository to each client for training, a pre-trained model based on OAI data is created at the server and sent to clients to reduce data traffic.
As shown as the 3D images in Fig.~\ref{fig:overview}, the knee cartilages are thin tissues, thus posing challenges to local training with limited supervision. Also, the image resolution of our private dataset (bottom) is significantly lower than the data repository (top), because coarse scanned images are more commonly used in clinical applications.
Such heterogeneity in resolution between the clinical dataset and the repository prohibits applying the model trained on OAI repository to the local dataset.

Instead of directly applying the pre-trained model, we can distill the knowledge of knee cartilages from the OAI repository to accelerate collaborative training. As illustrated in Fig.~\ref{fig:local}, our few-shot learning method contains a teacher-student architecture \cite{UA-MT} to distill the dual knowledge that consists of the response and feature-based \cite{gou2021knowledge} knowledge of the target from the pre-trained teacher model to the local student models. The response-based knowledge refers to the soft label created by the teacher network. The representation of knee cartilages produced by the encoder from the teacher network is used as feature-based knowledge. 
Then offline distillation is used to transfer the dual knowledge from the pre-trained model to the client-side model through unlabeled local data. The distillation process helps each client to extract a more general feature that is not bounded to the quality of data \cite{Nguyen:2022tx,mai2021few} and reduces the time and data transfer between clients and server through dual knowledge. In parallel, supervised learning adapts the collaborative model to the local dataset through the labeled data. 


In this paper, we propose a FL-based few-shot learning framework with dual knowledge distillation for improved segmentation of knee cartilages from 3D MR data. 
Our contributions are: (i) we identified the problem of limited local annotations among medical institutions and propose a few-shot learning method that utilize prior knowledge from well-annotated open data repository to train a collaborative deep learning model with few local annotations, (ii) we address the data heterogeneity problem of using Non-IID \cite{ZHU2021371} sources and a large disparity in imaging parameters between repository and local clinical data, and (iii) we identify the problem of massive data transfer associated with utilizing data repository in FL settings and solved it through prior feature extraction of the data repository.
Two state-of-the-art methods have been selected and compared to our method. Our method has shown superior performance.
\vspace{-0.2cm}
\section{Methods}
\begin{figure}
    \centering
    \includegraphics[width=0.90\linewidth]{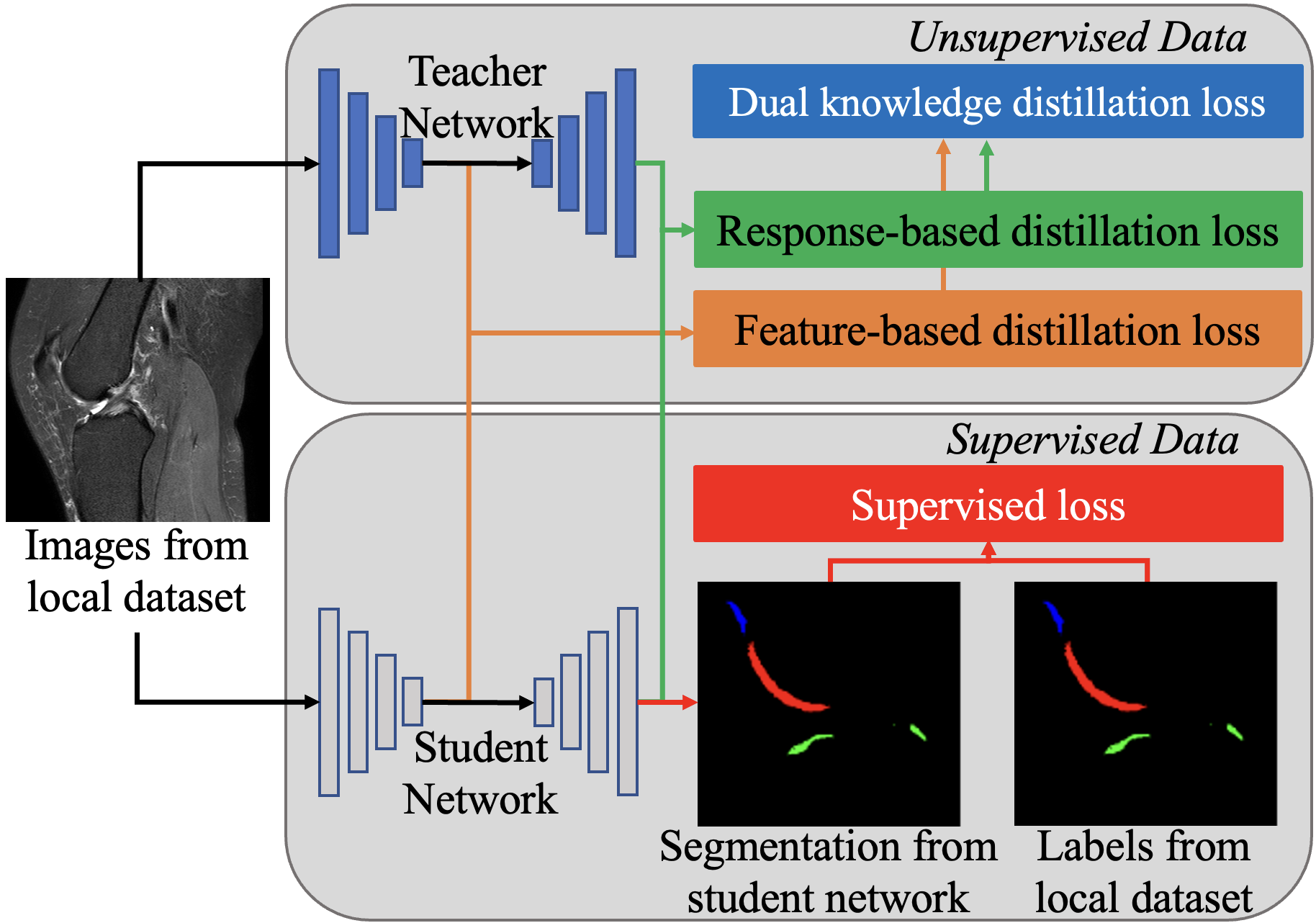}
    \caption{Our few-shot learning method with dual knowledge distillation in each client. The unsupervised images are used to gain response and feature-based knowledge from the teacher network and distill it to the student network. The student network is also optimized on the supervised images.}
    \label{fig:local}
\end{figure}
\label{sec:methods}


%
\textbf{Few-shot Learning with Dual Knowledge Distillation.} Fig.~\ref{fig:local} shows our client-side few-shot learning framework with supervised and dual knowledge distillation loss terms. The supervised loss consists of the cross-entropy loss and the dice loss between the segmentation of the student network $\mathbf{S}$ and the ground truth to provide feedback, defined as $L_S(\mathbf{S})$. 

Since the private dataset is heavily unlabeled, our method guides the training on unlabeled images in the support set by exploiting the dual knowledge acquired from the training set. This helps alleviate the label shortage during training of the student network with the local images. We define the dual knowledge distillation loss $L_{DKD}$ using the response-based $L_R$ and feature-based knowledge distillation loss $L_F$.

The response-based knowledge is extracted using the soft label produced by the teacher model $\mathbf{T}$. The soft label contains the probability distribution of each voxel. Thus, it contains more information compared to the binary-encoded hard label. However, the teacher may produce incorrect labels because the teacher model is trained on the OAI repository, which is different from the private data. Such incorrect segmentation needs to be identified and excluded. Inspired by \cite{UA-MT}, we estimate the uncertainty of the teacher-produced soft label by utilizing multiple scholastic passes via random dropout and adding noise to the unsupervised data to get multiple soft labels. Then predictive entropy on the soft labels is utilized to assess the uncertainty of the teacher network. The score is then used to filter unreliable predictions and select confident labels for the student to learn. This process constitutes the response-based distillation loss $L_{R}(\mathbf{T},\mathbf{S})$.

To further accelerate training and improve the accuracy of the collaborative model, feature-based knowledge of the teacher model is used. The idea is to capture the high-level representation of the target from the teacher network trained on the data repository. Since the teacher network is a pre-trained model, the feature maps of the teacher model are better than the randomly initialized student model. Therefore, the goal is to let the student network produce a similar set of feature vectors with the same cartilage as the teacher network to expedite the training process. 
With this in mind, we distill the feature-based knowledge by utilizing the KL divergence on the latent code produced by the encoder network in both the teacher and student networks. Let $\text{E}_{\mathbf{T}}$ and $\text{E}_{\mathbf{S}}$ be the encoder of teacher and student network, then $L_F(\mathbf{T},\mathbf{S})$, which is the feature-based distillation loss with input $x$, is:
\vspace{-0.2cm}
\begin{equation}
    L_F(\mathbf{T},\mathbf{S})=\sum_{j\in x}\text{E}_{\mathbf{T}}(j)\log{\frac{\text{E}_{\mathbf{T}}(j)}{\text{E}_{\mathbf{S}}(j)}}
    \vspace{-0.2cm}
\end{equation}
Therefore, the loss function $L$ of our few-shot learning is
\vspace{-0.2cm}
\begin{align}
    L(\mathbf{T},\mathbf{S})&=L_{S}(\mathbf{S})+\lambda L_{DKD}(\textbf{T},\textbf{S)} \\&= L_{S}(\mathbf{S})+\lambda(L_{R}(\mathbf{T},\mathbf{S})+L_{F}(\mathbf{T},\mathbf{S}))
    \vspace{-0.3cm}
\end{align}
where $L_S, L_{DKD}, L_R, L_F$ stands for supervised loss, dual knowledge distillation loss, response-based distillation loss, and feature-based distillation loss, respectively. $\lambda$ regularizes the supervised and dual knowledge distillation losses.

Although the teacher network can provide valuable insights in the first few rounds of training, the effectiveness of the knowledge distillation diminishes with the student model performing better on the private dataset. Eventually, the teacher model will hold back the performance of the student model. However, during the first few communication round, the student model barely contains any knowledge regarding the morphology of the cartilages. Updating the teacher network from the student network will undermine the accuracy of the teacher network. Thus, a delayed exponential moving average (EMA) update from student to teacher is applied.

\noindent\textbf{Inter-institutional Federated Learning.} To share the knowledge gained from each client, we integrate our proposed few-shot learning method into the federated learning framework. No patient data will be transferred in any part of the training process. In our paper, we facilitate the federated learning framework similarly to FedAVG \cite{FedAVG} as indicated in Fig.~\ref{fig:overview}. 
The federated learning process is outlined in Alg.~\ref{alg:fedavg}: Let $\mathbf{S}^c_t$ be the student model weights from $c\in C$ in the synchronization round $t$:
\begin{algorithm} 
	\caption{In the cluster, there are $N=|C|$ clients in total, each with a learning rate of $\alpha$. The set containing all clients is denoted as $C$. The communication interval is denoted as $E$.}
	\label{alg:fedavg} 
	\begin{algorithmic}[1] 
	    \item[\textbf{Central server do:}]
	        \State Initialize student model with random weights $\mathbf{S}_0$
	        \State Load and distribute the pre-trained teacher model $\mathbf{T}$
	        \For {each communication rounds $t \in {1, ..., rounds}$}
	            \ForAllP {each client $c \in C$}
	                \State $\mathbf{S}_{t}^{c} \leftarrow$ TrainLocally$(c, \mathbf{S}_t)$ \# Collect models
	            \EndForAllP
	            \State $\mathbf{S}_{t+1} \leftarrow \sum_{c=0}^{N} p_{c}\mathbf{S}^{c}_t$ \# Aggregate client models
	        \EndFor 
	\item[\textbf{TrainLocally($\mathbf{S}_0$):}]
	    \For {each client iteration $e \in {1, ..., E}$}
	        \State $\mathbf{S}_{e} \leftarrow \mathbf{S}_{e-1} - \eta \nabla L(\mathbf{T},\mathbf{S}_{e-1})$ \# Perform local training
	    \EndFor
	    \If {$t\geq 6$}
	        \State $\mathbf{T}\leftarrow $ UpdateEMA($\mathbf{T},\mathbf{S_E}$) \# Delayed EMA update
	    \EndIf
	    \State \Return $\mathbf{S}_{E}$
	\end{algorithmic} 
\end{algorithm}
\vspace{-0.5cm}
\section{Experiments}
\begin{table*}[t]
    \centering
    \begin{tabular}{c|c|c|c|c|c|c|c|c|c|c|c|c}
 & \multicolumn{3}{c|}{All Cartilages} & \multicolumn{3}{c|}{Femoral Cartilage} & \multicolumn{3}{c|}{Tibial Cartilage} & \multicolumn{3}{c}{Patellar Cartilage}\\
 \cline{2-13}
 & DSC & VOE & ASSD & DSC & VOE & ASSD & DSC & VOE & ASSD & DSC & VOE & ASSD\\
  \hline
  Local & 0.713 & 43.981 & 1.228 & 0.727 & 42.193 & 1.297 & 0.710 & 44.647 & 0.822 & 0.566 & 58.029 & 2.646\\
    \hline
Semi & 0.744 & 40.199 & 1.182 & 0.767 & 37.397 & 1.287 & 0.757 & 38.803 & 0.794 & 0.622 & 52.560 & 2.367\\
\hline
SSFL & 0.746 & 39.973 & 1.100 & 0.740 & 40.344 & 1.608 & 0.754 & 39.236 & 0.631 & 0.654 & 50.152 & 1.464\\
\hline
Fed-Semi & 0.762 & 38.310 & 0.902 & 0.763 & 38.125 & 1.022 & 0.756 & 39.063 & 0.629 & 0.673 & 48.158 & 1.500\\
\hline
Ours & \textbf{0.789} & \textbf{34.529} & \textbf{0.643} & \textbf{0.796} & \textbf{33.386} & \textbf{0.632} & \textbf{0.777} & \textbf{36.309} & \textbf{0.523} & \textbf{0.690} & \textbf{45.464} & \textbf{1.441}
    \end{tabular}
    \caption{Quantitative comparison of methods on our private dataset. The best results have been highlighted in the chart.}
    \label{tab:Comparison_btw_methods}
\end{table*}
\begin{figure*}
    \centering
    \includegraphics[width=0.83\linewidth]{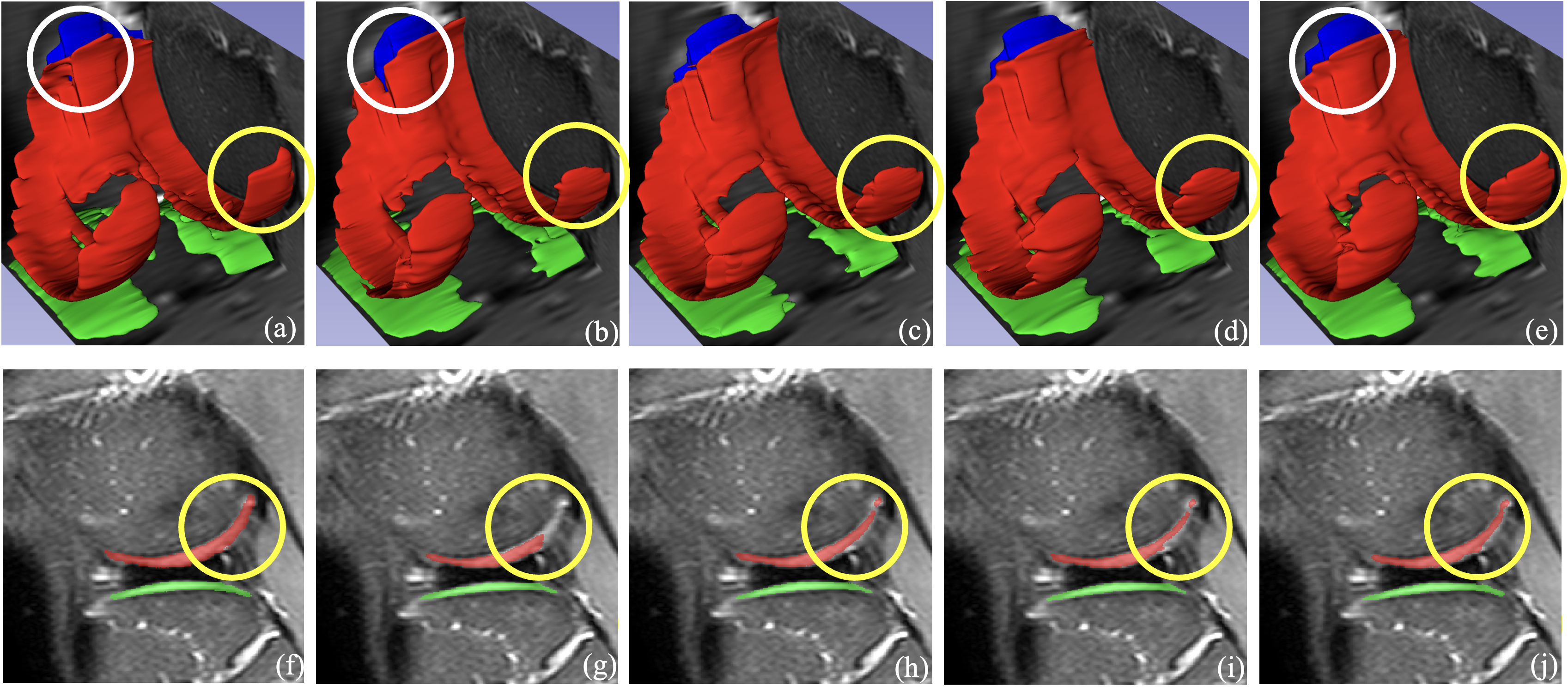}
    \caption{Visual results of subject 1. (a) and (f) shows the GT labels; (b) and (g) are from local training; (c) and (h) are from SSFL; (d) and (i) are Fed-Semi; (e) and (j) are from our proposed method in 3D and sagittal views, respectively.}
    \label{fig:10006}
    \vspace{-0.5cm}
\end{figure*}


\textbf{Experiment settings.} We evaluate our method on a private dataset, which contains $20$ labeled and $1000$ unlabeled 3D MR knee images in each of the 4 clients. The voxel size (mm) of the images ranges from $(0.303, 0.303, 3.5)$ to $(0.3125, 0.3125, 4.5)$. 
The dual knowledge is extracted from the OAI repository with voxel size of $(0.365,0.365,0.7)$. 
All images are resized to $352\times 288\times 16$, and their pixel intensity has been normalized to $[0,1]$. All datasets have been split into $6:2:2$ for training, validation, and testing. We utilize U-net \cite{HeUnet} as the segmentation network. 

Two state-of-the-art federated semi-supervised segmentation approaches, SSFL \cite{SSFL} and Fed-Semi \cite{FedSemiSupervised}, are compared. We also evaluated the performance without federated learning (Local) and without knowledge distillation (Semi). To measure accuracy and spatial correctness, dice similarity coefficient (DSC), volumetric overlap error (VOE) (mm$^3$), and average symmetric surface distance (ASSD) (mm) between the GT labels and segmentation are reported.
\begin{figure}[t]
    \centering
    \includegraphics[width=0.70\linewidth]{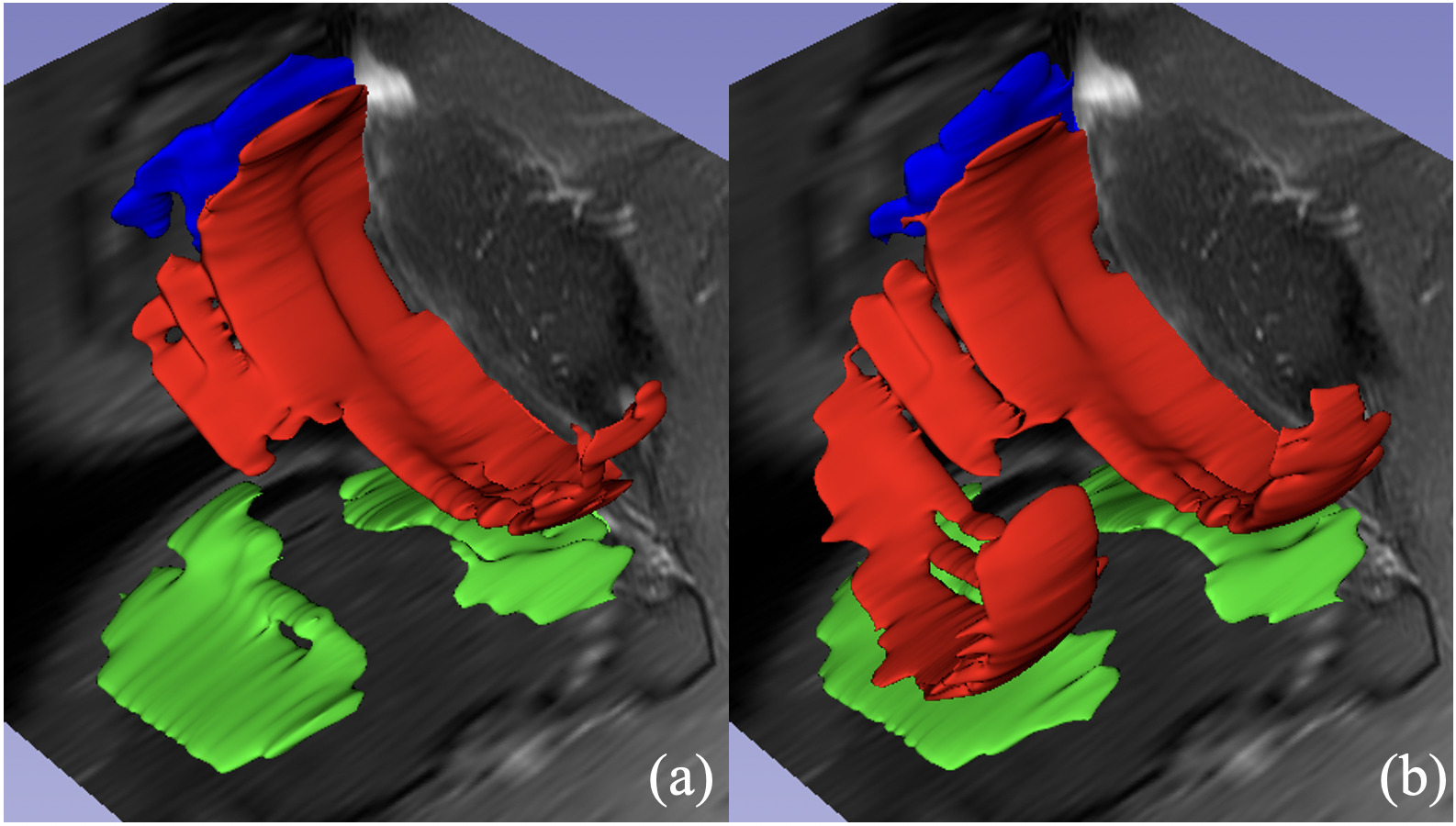}
    \caption{Visual result for subject 2. (a) SSFL and (b) Ours.}
    \label{fig:hc}
    \vspace{-0.5cm}
\end{figure}

\noindent \textbf{Experiment results.} As observed in Table~\ref{tab:Comparison_btw_methods}, our method outperforms both SSFL and Fed-Semi, with a DSC increase of $5.7\%$ and $3.6\%$, respectively.  
This indicates that the additional knowledge improved the training in the FL scenario. Furthermore, our 3D surface is more accurate, since the ASSD is reduced by $41\%$. By comparing our method to local training, federated learning improves DSC by $10.6\%$ for all cartilages. Prior knowledge is advantageous in small target segmentation since it provides additional information about it. For example, PC is a small cartilage, and segmenting PC is challenging since there are fewer voxels representing that cartilage. Our method has shown an increase of $11\%$ in DSC score on PC compared to those without prior knowledge. For efficiency, our method utilizes 19 communication rounds compared to 28 rounds needed by other methods, which amounts to a $32\%$ decrease in data transfer and training time. 

Fig. \ref{fig:10006} shows three cartilages of one subject with the above-mentioned methods. Comparing the white circled region, local training (b) under segments PC compared to our method (e), which confirms that the dual knowledge helps the small cartilage segmentation. Meanwhile, as indicated in the yellow circled area of Fig. \ref{fig:10006}, our method produced the most accurate result compared to other methods, which all failed to discover parts of FC. In particular, about one-third of FC produced by local training (g) is missing. This provides evidence that clients with limited labels cannot train a usable network, and collaboration between medical institutions is needed. In addition, both SSFL (h) and Fed-Semi (i) methods show discontinuous labels of FC indicating insufficient knowledge of cartilage shape compared to the proposed method. 
To show the stability of our method, Fig.~\ref{fig:hc} demonstrates a hard case in our dataset. The label produced by SSFL missed a significant portion of FC compared to our method, which is not acceptable for medical applications. However, our method maintains high accuracy throughout the test cases because of the additional knowledge distilled from the OAI repository.  
\vspace{-0.2cm}
\section{Conclusion}
\label{sec:conclusion}
In this work, we proposed a few-shot FL framework with dual knowledge distillation. 
The dual knowledge, including the response and feature-based knowledge extracted from the data repository on the server side, is used to accelerate and guide the training of the student model locally using the private dataset. Our few-shot learning reduces annotation requirements in each client, and knowledge distillation mitigates the challenge of dissimilarity in imaging resolution and parameters of the training and support set. We carried out a comprehensive analysis of our method and obtained superior results. 
\section{Compliance with Ethical Standards}
\label{sec:CES}
This research study was conducted retrospectively using human subject data including open access dataset by National Institution of Health through the Osteoarthritis Initiative and private dataset from Shanghai Sixth People's Hospital. The studies involving human participants were reviewed and approved by Ethics Committee of Shanghai Sixth People's Hospital. Written informed consent to participate in this study was provided by the participants’ legal guardian/next of kin. 
\bibliographystyle{IEEEbib}
\bibliography{refs.bib}

\end{document}